\documentclass[amsmath,12pt,amssymb,preprint,prd,aps,nofootinbib]{revtex4}
\usepackage{amsfonts} %\usepackage{citesort}
\usepackage{graphicx} % Include figure files
\usepackage{epsfig}
\usepackage{multirow}
\usepackage{bm}% bold math
\usepackage{subfig}
\usepackage[dvipsnames]{xcolor}
\begin{document}
\title{Compressibility and speed of sound in magnetized nuclear matter with broken scale invariance}

\author{Pallabi Parui$^{1}$}
\email{pallabiparui123@gmail.com}
\author{Nilanjan Chaudhuri$^{2}$}
\email{nilanjan.vecc@gmail.com; n.chaudhri@vecc.gov.in}
\author{Pradip Roy$^{1,3}$}
\email{pradipk.roy@saha.ac.in} 
\author{Sourav Sarkar$^{2,3}$}
\email{sourav@vecc.gov.in}
\affiliation{$^{1}$Saha Institute of Nuclear Physics, 1/AF Bidhannagar, Kolkata - 700064, India}
\affiliation{$^{2}$Variable Energy Cyclotron Centre, 1/AF Bidhannagar, Kolkata - 700064, India}
\affiliation{$^{3}$Homi Bhabha National Institute, Training School Complex, Anushaktinagar, Mumbai - 400085, India  
}
\begin{abstract}
The thermodynamical properties of magnetized nuclear matter at finite temperature and baryon chemical potential are studied within an effective model incorporating the QCD trace anomaly effect. The presence of magnetic field induces anisotropic structure in the energy momentum tensor due to the broken rotational invariance. The study exhibits a phase transition through the sudden change of the effective nucleon mass in a certain range of baryon chemical potential and temperature. The addition of nucleonic vacuum contribution at finite magnetic field leads to the magnetic catalysis effect. The change in squared speed of sound with chemical potential at various temperatures is closely connected to the nature of phase transition in nuclear matter. The pressure anisotropy results in different values of sound speed and isothermal compressibility in the parallel and perpendicular directions with respect to the magnetic field. The smaller values of isothermal compressibility in the parallel direction compared to the perpendicular one indicate that the equation of state is stiffer along the magnetic field direction. The studies of these thermodynamic observables can have significant importance in analyzing the properties of some compact astrophysical objects as well as in the context of non-central heavy ion collision experiments.   
\end{abstract}
\maketitle
\section{Introduction}
\label{sec1}
The study of the characteristic properties of hot and/or dense strongly interacting matter is an ongoing topic of research in the context of relativistic heavy-ion collision experiments and some compact astrophysical objects. Investigation of the thermodynamic properties in a background magnetic field has gained much interest due to the induced anisotropic behavior in some important thermodynamic observables for both the quark matter and nuclear matter studies. In the off-central ultra-relativistic heavy ion collision experiments, strong magnetic field has been estimated at the initial stages of the collision due to the highly energetic, oppositely directed motion of two colliding ion beams at RHIC, BNL and LHC, CERN \cite{sokov, deng, tuchin}. The generated fields rapidly decay within a few fm/c. However, the decay process might be delayed sufficiently due to the non-zero electrical conductivity of the produced medium which may sustain a finite magnetic field even in the subsequent hadronic stages following the initial plasma phase \cite{gursoy, inghirami}. Important physical phenomena arise due to the effects of the magnetic field on the QCD vacuum properties e.g., magnetic catalysis \cite{kharzeevmc, shovkovy, gusy}, inverse magnetic catalysis \cite{Preis}, chiral magnetic effect \cite{kharzeev, fukushima}, etc. A reasonably strong magnetic field has been inferred in some compact astrophysical objects. The effects of the magnetic field have been studied extensively on the equation of state (EOS) and structure of neutron star matter e.g., magnetars, proto-neutron star, etc. \cite{rabhi, brodrick, wei, mao}. The strength of the magnetic field in the compact stellar objects vary significantly from the inner core to the surface area. \textcolor{black}{From the scalar virial theorem based on Newtonian gravity, it is assumed that the interior field strength in neutron stars can be as large as $\sim 10^{18}$ G \cite{shapiro, dong}. Also, the matter density in the neutron star core may exceed up to several orders of the nuclear matter density \cite{debadesh}.} %The estimated field strengths in the inner core can lie in the range of $10^{18}$ G \cite{dong} to $10^{20}$ G \cite{pressure1} for different stellar systems. These large inner core fields in the magnetars may appear due to the conservation of magnetic flux in stellar media of very high electric conductivities, 
The surface magnetic fields of magnetars and pulsars can be of the order of $10^{14}-10^{15}$ G \cite{vasisht, tiengo} and $10^{8}-10^{12}$ G \cite{livingstone}, respectively.% There have been spectroscopic and spin-down studies of the soft-gamma ray repeaters and anomalous x-ray pulsars in deduction of these surface magnetic fields. %The strength of the magnetic fields produced are of the order of QCD scale parameter $|eB|\approx \Lambda_{QCD}^2$, thus can give considerable observable effects of QCD.

The in-medium behavior of some thermodynamic observables e.g. compressibility, speed of sound etc., are correlated with the properties of dense matter system such as the phase structure and  stiffness of the equation of state (EOS). The EOS is an important quantity in characterizing the compact star objects \cite{prakash}. It has been illustrated that a quark star is self-bound by the strong interaction and neutron star is gravitationally bound \cite{star1, star2}. The fundamental properties of mass, radius, moment of inertia, etc., of the magnetized neutron stars are sensitive to the effects of magnetic field on the EOS at the inner core of the star. The compressibility of matter act as an important parameter to study the stiffness of the EOS. %and analysis of the effects of magnetic field on its behavior. %There exist two distinct concepts of incompressibility (compression modulus) and compressibility to indicate the relative stiffening of the EOS. The compression modulus measures the ability to withstand compression and the compressibility reflects its opposite behavior.
It has been studied for both the nuclear and quark matters \cite{blaze1, dexhe}. The Nambu-Jona-Lasinio (NJL) model and Polyakov-loop extended NJL (PNJL) model have been used to study the compressibility of quark matter at zero \cite{avijit, compressibility2} and finite \cite{prd110, compressibility1} values of magnetic field. %An increased compressibility has been observed in the chiral symmetry broken phase which eventually diverges near the tricritical point \cite{compressibility2}. Its in-medium behavior is, therefore, sensitive near the phase transition regions. 
Another important thermodynamic quantity is the speed of sound ($c_s$) which measures the propagation of a compression wave through the medium. It is derived from the change in thermodynamic pressure with respect to a change in the energy density for a fixed parameter corresponding to the different profiles of EOS \cite{prc107ss}. It act as a significant probe for the EOS in studying the vast area ranging from the space-time evolution of the generated strongly interacting matter in heavy-ion collisions to the investigation of the neutron star properties. % The dependence of $c_s$ on the medium effects of density (chemical potential), temperature, etc., may provide useful information in the description of fireball evolution and final state observables. 
In the study of the QCD phase structure a relation between the sound speed and baryon number cumulants has been developed in ref.\cite{sorensen}. % The physical quantity of $c_s$ may provide valuable information on the characteristics of the interior of stellar matter, including the stiffness of the EOS, and the maximum possible mass of a neutron star at a given radius. A non-monotonic behavior in the sound speed has been observed as a function of density (from hadron to free quark phases) in a magnetized hybrid neutron star \cite{pressure2}. The anisotropic nature of $c_s$ due to the pressure anisotropy in a magnetized system has been studied from the first principle calculations. %The variation of $c^2_s$ as a function of density may have important consequences on the star's mass-radius relation, tidal deformability, acting as a sensitive probe of the EOS of neutron star matter and the quark-hadron phase transition in its high density core.
%The astrophysical constraints on the masses and radii of neutron star as obtained from the analysis of X-ray observations put constrain on the speed of sound in the dense core, which leads to exceed the conformal limit of $c_s^2\leq 1/3$ at larger densities compared to the nuclear saturation density \cite{tews}.
In the context of neutron star research, the in-medium behavior of $c_s$ as a function of density has considerable impact on the star's mass-radius relation, its cooling rate, tidal deformability, etc. Some studies show that, for a massive neutron star matter it is necessary to have a density range in which the EOS becomes very stiff and the squared speed of sound becomes much larger than $1/3$ \cite{reed, tidal, tews, reddy, holographic}. The sound speed is shown to be an important quantity for the gravitational wave frequencies induced by the $g$ mode oscillation of a neutron star \cite{gmode}.  %The speed of sound exceeding the conformal limit has been observed in several other works for e.g., in holographic approaches \cite{holographic}, chiral effective field theory \cite{tews}, quarkyonic matter \cite{reddy}, etc. 
The effects of chemical potential, density, temperature and/or magnetic field on $c_s$ have been studied using a number of methods: lattice QCD \cite{lattice}, Nambu-Jona-Lasinio (NJL) and Polyakov-Nambu-Jona-Lasinio (PNJL) models \cite{pnjl1,pnjl2}, quark-meson coupling model \cite{qmcss}, the hadron resonance gas model \cite{hrgss}, etc. The in-medium behavior of the speed of sound has been investigated in the full QCD phase diagram spanning from the QCD matter \cite{prd105ss} to nuclear matter regimes \cite{prc107ss}. %The values near the transition regions are estimated with a limiting value of $1/3$ after chiral restoration at high temperature. In-medium adiabatic sound speed sheds light on the important characteristics of the QCD phase structures near the critical end point and crossover regions. %In the nuclear matter study of the sound speed, the thermodynamic quantities needed to compute $c_s^2$ for different fixed parameters have been calculated from the Lagrangian of the non-linear Walecka model \cite{prc107ss, prc109}. The thermodynamic potential of the grand canonical ensemble is written under the mean-field approximation and the pressure and energy density are obtained from the thermodynamic relations of the ensemble. Therefore, the sound speed is derived for a specific fixed quantity for the propagation of the compression wave through the medium. As there is close connection between the nature of nuclear liquid-gas (LG) transition and the chiral first-order phase transition, the study of $c_s$ within the nuclear matter environment can be compared with its behavior in quark matter and therefore important conclusions can be derived in the context of phase transitions and the nature of EOS.  
The presence of magnetic field leads to the breaking of the $O(3)$ rotational symmetry leading to an anisotropic structure in the energy momentum tensor of the system. It results in pressure anisotropy giving rise to different pressures in the longitudinal and transverse directions with respect to the magnetic field \cite{pressure1, pressure2, monika}. This effect further results in the anisotropic behavior of the compressibility and speed of sound. There may arise instabilities related to the pressure anisotropy which appear as its derivative (in either the longitudinal or transverse direction) changes sign with respect to density \cite{pressure1, instability2, instability3, instability4}. %These instabilities have been considered for the magnetized fermionic matter \cite{pressure1} and quark matter \cite{instability2} due to the vanishing of the parallel pressure and for the electron gas \cite{instability3}, strange quark matter \cite{instability4} due to the vanishing of perpendicular pressure. 

%The medium effects may have significant impact on the equation of state of the hot and/or dense magnetized matter. This can further affect a number of thermodynamic observables including the compressibility and speed of sound which are correlated with the characteristics properties of dense matter system such as the phase structure and stiffness of the EOS.
 %Furthermore, the cooling processes, magnetic field evolution should have strong dependency on the properties of matter under high magnetic fields. In \cite{prakash}, the EOS of a multi-component interacting matter has been studied under a strong magnetic field by applying a field theoretical approach in the context of neutron star matter. The effects of Landau quantization and anomalous magnetic moments of nucleons lead to the softening and stiffening of the EOS, respectively, at very strong magnetic fields.  

\textcolor{black}{In literature, there are several effective model Lagrangians with baryonic and mesonic degrees of freedom to study the non-perturbative regime of QCD at high baryon density and/or high temperature.} %  thermodynamic properties of nuclear matter \cite{glendenning, boguta}. 
In the Walecka model Lagrangian, the attractive and repulsive nature of the scalar meson-nucleon ($\sigma-N$) and vector meson-nucleon ($\omega-N$) interactions provide the minimum of the binding energy at the saturation point of nuclear matter with the phenomenologically fitted model parameters at saturation \cite{glendenning}. 
%The model parameters are fitted  to the nuclear matter saturation properties. %: the saturation density ($\rho_0=0.153fm^{-3}$), binding energy per nucleon ($-16.3$ MeV) etc.
In order to reproduce the empirical values of compression modulus near saturation, the Lagrangian has been extended to incorporate the non-linear scalar self-interactions \cite{boguta}. %The scalar field equation is solved self-consistently under the mean-field approximation to find the in-medium effective nucleon mass and other thermodynamic quantities including energy density, pressure, compression modulus, symmetry energy of the infinite nuclear matter system. 
\textcolor{black}{There are also some effective model studies on QCD phase transition by incorporating the concepts of QCD chiral and scale symmetries in the description of hadronic interactions with the quark and gluon condensates \cite{ellis273}. The medium modifications of these condensates have been studied in nuclear matter to explore the expected phase boundary between the nuclear matter and quark-gluon phase at high density and/or high temperature regimes of phase diagram \cite{ellis273}, which can be explored in heavy ion collision experiments. High temperature transitions are relevant for the cosmological studies whereas, the high baryon-density ones are most convenient in the neutron star studies. The condensates act as the order parameters of phase transition.} 
In ref.\cite{heide293}, a scalar glueball potential has been introduced to account for the QCD trace anomaly effect in a chiral Lagrangian. This leads to an extended range of the possible solutions with improved shape of the nuclear saturation curve near equilibrium. 
\textcolor{black}{In ref.\cite{gelman}, the hadronic properties of pion-nucleon scattering and meson masses etc., have been studied using the linear $\sigma$ model approach by accounting for the basic symmetry principles and scaling of QCD.
The linear $\sigma$ model of Gell-Mann and Levy has been extended in ref.\cite{papa55} to incorporate the chiral and scale symmetry breaking effects of QCD within an effective model approach. %An appropriate dimension of the scaled scalar, isoscalar dilaton field $\chi$, i.e., $\chi/\chi_0$ is multiplied to the chiral invariant potential $\sim (\sigma^2+\pi^2)$ for scale invariant. 
The logarithmic potential in terms of the scaled dilaton field $\chi/\chi_0$, accounts for the scale-invariance breaking effect of QCD and thus producing the non-zero trace of the energy momentum tensor. The finite value for the dilaton field leads to the spontaneous chiral symmetry breaking of the Lagrangian. Furthermore, the chiral symmetry is broken explicitly which gives rise to the pion mass. The vector meson mass is generated dynamically by the $\sigma$ and $\chi$ fields by adjusting a model parameter appropriately \cite{papa55}.}

The nuclear matter EOS including the composite nature of nucleons has been studied at finite temperature within the quark meson coupling (QMC) model with quarks coupled to the scalar and vector mesons \cite{qmcpanda}. The nuclear EOS, entropy density, and effective nucleon mass have been calculated within this QMC model as a function of baryon density at different temperatures by considering medium dependent bag parameter. %The effects of the scalar, isovector field $\delta$ have been studied in the framework of relativistic mean-field theory of nuclear matter in addition to the $\sigma$, $\omega$ and $\rho$ mesons constituting nuclear interactions \cite{wei}. The effects have been considered on the effective mass of nucleon and EOS of the neutron star matter at finite magnetic field accounting for the non-zero anomalous magnetic moments of the nucleons.
\textcolor{black}{The relativistic mean-field model of nuclear matter % where the constituents protons and neutrons interacting via the exchange of scalar isoscalar $\sigma$, vector isoscalar $\omega$ and vector isovector $\rho$ mesons,
has been used extensively in the literature to study the properties of dense hadronic matter e.g., the structure of neutron star at finite magnetic field \cite{mao, wei}. % The nuclear interactions are described through the exchange of the scalar isoscalar $\sigma$, vector isoscalar $\omega$ and vector isovector $\rho$ mesons \cite{mao} as well as a scalar isovector field $\delta$ \cite{wei}. 
The standard relativistic mean-field model is extended to account for the QCD trace anomaly condition in the study of finite density matter \cite{kapusta44}. In our present work, we have incorporated the effects of finite temperature and magnetic field within the framework of the relativistic mean-field Lagrangian incorporating the scale-invariance breaking gluonium potential. The effect of the scale-symmetry is considered by adding a kinetic energy in the scalar gluonium or dilaton field $\chi$ and scaling the mass terms with the appropriate power of the scaled field ($\chi/\chi_0$). The scale-symmetry breaking logarithmic potential in terms of $\chi/\chi_0$ leads to the non-zero trace of the energy momentum tensor. The model parameters have similarly been fitted using the empirically known saturation properties of nuclear matter at zero temperature and zero magnetic field. Here we only consider the symmetric nuclear matter system as an initial step towards studying the thermodynamics of the system.}% However, in the future proceeding work these properties are planned to investigate by incorporating the effects of isospin asymmetry within an effective hadronic model approach.}
%The thermodynamic potential of a grand canonical ensemble is thus used to calculate the thermodynamic observables at finite  chemical potential, temperature and magnetic field. 

The present paper is organized as follows: in sec.\ref{sec2}, the scale-invariance breaking effective Lagrangian is briefly discussed to find the in-medium thermodynamic quantities of the magnetized symmetric nuclear matter system at finite temperature. In sec.\ref{sec3}, the basic thermodynamic relations are obtained from the thermodynamic potential of a grand canonical ensemble to calculate the important properties of magnetization, compressibility and speed of sound in the matter. The results of the present investigation are discussed in sec.\ref{sec4}. Finally, sec.\ref{sec5}, summarizes the findings of the present work.  

\section{The effective model Lagrangian}
\label{sec2}
The present model consists of nucleons, scalar and vector mesons and the scalar gluonium field as the effective degrees of freedom in the Lagrangian. It has been used to study isospin symmetric nuclear matter at finite density. The general structure of the Lagrangian incorporating the kinetic energy and interaction terms of the nucleons ($\psi$), scalar isoscalar meson $\sigma$, vector isovector meson $\omega^{\mu}$ and the scalar gluonium field $\chi$, is \cite{kapusta44}
\begin{multline}
    \mathcal{L}= \bar{\psi}(i\gamma_{\mu}D^{\mu}-M_N\frac{\chi}{\chi_0}+g_{\sigma}\sigma-g_{\omega}\gamma_{\mu}\omega^{\mu})\psi - \frac{1}{4}\omega_{\mu\nu}\omega^{\mu\nu} +\frac{1}{2}m_{\omega}^2\left(\frac{\chi}{\chi_0}\right)^2\omega_{\mu}\omega^{\mu} \\ + \frac{1}{2}(\partial_{\mu}\sigma)(\partial^{\mu}\sigma) - \frac{1}{2}m_{\sigma}^2\left(\frac{\chi}{\chi_0}\right)^2\sigma^2 -\frac{1}{3}b m_N \frac{\chi}{\chi_0} (g_{\sigma}\sigma )^3 -\frac{c}{4}(g_{\sigma}\sigma )^4 \\ + \frac{1}{2}(\partial_{\mu}\chi)(\partial^{\mu}\chi) -B_g\left[4\left(\frac{\chi}{\chi_0}\right)^4ln\left(\frac{\chi}{\chi_0}\right) - \left(\frac{\chi}{\chi_0}\right)^4 +1 \right] -\frac{1}{4}F_{\mu\nu}F^{\mu\nu},
    \label{lag1}
\end{multline}
where, $D^{\mu}=(\partial^{\mu}+ieA^{\mu})$ is the covariant derivative, replacing the nucleon kinetic energy in the first term to represent the charged particle interaction in a homogeneous background magnetic field in the $z$ direction given by $A^{\mu}=(0,yB,0,0)$. $\psi$ represents the nucleon doublet in the isospin space; with $e$, the electric charge of the proton ($p$). In eq.(\ref{lag1}), $\omega^{\mu\nu}=(\partial^{\mu}\omega^{\nu}-\partial^{\nu}\omega^{\mu})$ and $F^{\mu\nu}=(\partial^{\mu}A^{\nu}-\partial^{\nu}A^{\mu})$, are the field strength tensors corresponding to the vector-meson field $\omega^{\mu}$ and the electromagnetic field $A^{\mu}$, respectively. The attractive and repulsive contributions of the nuclear force are included in a conventional way through the exchanges of the scalar $\sigma$ and vector $\omega^{\mu}$ mesons between the nucleons, respectively. The Lagrangian is made scale-invariant by multiplying the mass terms through the appropriate dimension of $(\frac{\chi}{\chi_0})$ and incorporating the kinetic energy of the gluon field $\chi$. The logarithmic scalar-gluon potential \cite{sechter21} 
\begin{equation}
    U(\chi)=B_g\left[4\left(\frac{\chi}{\chi_0}\right)^4ln\left(\frac{\chi}{\chi_0}\right)- \left(\frac{\chi}{\chi_0}\right)^4 +1 \right]
    \label{gluonpot}
\end{equation}
implements the effects of QCD trace-anomaly into the effective model Lagrangian. The minimum of $U(\chi)$ at $\chi=\chi_0$ gives rise to a gluon condensate, where the fluctuations around this minimum are referred to as the glueballs of mass $m_{glue}=\frac{4B_g^{1/2}}{\chi_0}$ \cite{kapusta44}. The constant $B_g$ gives the difference between the energy densities in the true non-perturbative vacuum with $\chi=\chi_0$ and perturbative one with $\chi=0$. Thus it can be identified with the constant factor in the MIT Bag model \cite{mitbag}. The scale-invariance breaking effects lead to the non-zero trace of the QCD energy momentum tensor. It can be expressed as 
$T_{\mu}^{\mu}=\langle\frac{\beta}{2g}G^{a\mu\nu}G^{a}_{\mu\nu}\rangle$, which contains the QCD renormalization group function $\beta$ and gluon field strength tensor $G^{\mu\nu}$. The $\beta$ function vanishes at the classical level, which suggests that scale-invariance breaking is included by the quantum effects.\\ To solve for the equations of motion as derived from the Lagrangian (\ref{lag1}), mean field approximation (MFA) is adopted. In this approximation, the meson fields are treated to be classical and the fluctuations around their classical average values have been neglected. The effective mass and chemical potential of the nucleon, under this approximation are: 
\begin{align}
    m^*_N&=m_N\frac{\chi}{\chi_0}-g_{\sigma}\sigma,
    \label{effecmass}\\
    \mu^*&=\mu_B-g_{\omega}\omega.
    \label{effecpot}
\end{align}
respectively. Here, $m_N$ is the vacuum mass and $\mu_B$ is the baryon chemical potential of the nucleon. The average meson fields are used in these relations, where $\omega$ represents the temporal part of the vector meson field $\omega^{\mu}$ with vanishing spatial components under MFA. The mass-scaled meson-nucleon coupling parameters $g_{\sigma}/m_{\sigma}$, $g_{\omega}/m_{\omega}$ along with three other independent parameters $b$, $c$ and $B_g$ are fitted to reproduce the nuclear matter saturation properties. It includes pressure $P=0$, saturation density $\rho_0=0.153\ fm^{-3}$, binding energy $-16.3$ MeV, the Landau mass $\sqrt{m^{*2}_N+k_F^2}=0.83m_N$, and the compressibility $K\approx 300$ MeV at the saturation point \cite{kapusta348}. The values of the parameters thus used are: $g_{\sigma}/m_{\sigma}=0.0132\ \text{MeV}^{-1}$, $g_{\omega}/m_{\omega}=0.0105\ \text{MeV}^{-1}$, $b = 0.01$, $c = 0.0146$, $B_g = (200\ \text{MeV})^4$.

  %  \frac{g_{\sigma}}{m_{\sigma}}=0.0132 MeV^{-1}; \ \frac{g_{\omega}}{m_{\omega}}= 0.0105\ MeV^{-1}; \ b = 0.01 ; c = 0.0146; B_g = (200\ MeV)^4.

The thermodynamic potential at finite temperature and magnetic field is written in terms of the mean meson fields as obtained from the Lagrangian (\ref{lag1}), 
\begin{multline}
    \Omega =  \frac{1}{2}m_{\sigma}^2\left(\frac{\chi}{\chi_0}\right)^2\sigma^2 +\frac{1}{3}b m_N \frac{\chi}{\chi_0} (g_{\sigma}\sigma )^3 +\frac{c}{4}(g_{\sigma}\sigma )^4 + U(\chi) - \frac{1}{2}m_{\omega}^2\left(\frac{\chi}{\chi_0}\right)^2\omega_{0}^2  \\  + 2\beta^{-1}\int\frac{d^3k}{(2\pi)^3}[ln(1-f_n^+)+ln(1-f_n^-)]+ \beta^{-1}\frac{|eB|}{2\pi}\sum_{\nu=0}^{\infty}a_{\nu}\int \frac{dk_z}{2\pi}[ln(1-f_{p,\nu}^+)+ln(1-f_{p,\nu}^-)] \\ + \Omega^{mag}_{vac} +\frac{B^2}{2} \quad \quad \quad \quad \quad \quad \quad \quad \quad  
    \label{thermopot}
\end{multline}
The second line contains the contribution of the thermomagnetic matter part with constituents protons ($N=p$) and neutrons ($N=n$). Here, $f_{N(=p,n)}^{+}(E^N-\mu^*)\ [f_{N(=p,n)}^-(E^N+\mu^*)]$ denotes the fermion [antifermion] distribution function with $f(x)= \frac{1}{e^{x/T}+1}$ and $\beta=1/T$. The energy of the charge neutral neutron is $E^{(N=n)}(k)=\sqrt{k^2+m^{*2}_N}$. The energy levels of the electrically charged protons are quantized due to the Landau level contributions in the presence of a background magnetic field, $E^{(N=p)}_{\nu} (k_z)=\sqrt{k_z^2+m^{*2}_N+2\nu|eB|}$, for $\frac{1}{2}$ spin charged fermions with electric charge $e$. The term $a_{\nu}=(2-\delta_{\nu 0})$ denotes the spin degeneracy of each Landau level $\nu$. Here, the free magnetic contribution is given by the last term $B^2/2$. The effects of the magnetized vacuum are considered for the case of charged fermions through \cite{haber, prc109, menezes} 
\begin{equation}
    \Omega^{mag}_{vac} = -\frac{(|eB|)^2}{2\pi^2}\left[\zeta'(-1,y)+\frac{y^2}{4}+\frac{y}{2}(1-y)lny \right].
\end{equation}
Here, $y=\frac{m^{*2}_N}{2|eB|}$; $\zeta(x,y)=\sum_{i=0}^{\infty}\frac{1}{(i+y)^x}$ is the Hurwitz zeta function with its derivative $\frac{d\zeta(x,y)}{dx}|_{x=-1}=\zeta'(-1,y)$. \\
Minimization of the thermodynamic potential as 
\begin{equation}
    \frac{\partial\Omega}{\partial\sigma}=\frac{\partial\Omega}{\partial\omega}=\frac{\partial\Omega}{\partial\chi}=0,
\end{equation}
leads to the following coupled, non-linear equations of motion for the mean meson fields 
\begin{align}
     m_{\omega}^2\left(\frac{\chi}{\chi_0}\right)^2\omega_0 &=g_{\omega}\rho_B;
     \label{eom1}\\
      m_{\sigma}^2\left(\frac{\chi}{\chi_0}\right)^2\sigma +bm_N\frac{\chi}{\chi_0}g_{\sigma}^3\sigma^2 + cg_{\sigma}^4\sigma^3  &=- \frac{|eB|}{2\pi^2}m^*_N\left[y(1-lny)+\frac{1}{2}ln\left(\frac{y}{2\pi}\right)+ln\Gamma(y) \right]+ g_{\sigma}\rho^s; 
      \label{eom2}\\
     m_{\omega}^2\left(\frac{\chi}{\chi_0}\right)^2\omega_0^2- m_{\sigma}^2\left(\frac{\chi}{\chi_0}\right)^2\sigma^2 &=16B_g\left(\frac{\chi}{\chi_0}\right)^4ln\left(\frac{\chi}{\chi_0}\right) + m_N\frac{\chi}{\chi_0}\rho^s+\frac{b}{3}m_N\frac{\chi}{\chi_0}(g_{\sigma}\sigma)^3.
     \label{eom3}
\end{align}
In eqs.(\ref{eom1})-(\ref{eom3}), the baryon number density $\rho_B$ and the scalar density $\rho^s$ for protons and neutrons in nuclear matter at finite temperature and magnetic field can be expressed as 
\begin{align}
   \rho_B=\rho_n+\rho_p &=2\int\frac{d^3k}{(2\pi)^3}(f^+_n-f^-_n) + \frac{|eB|}{2\pi}\sum_{\nu=0}^{\infty}a_{\nu}\int\frac{dk_z}{2\pi}(f^+_{p,\nu}-f^-_{p,\nu}); 
   \label{noden}\\
\rho^s=\rho^s_n+\rho^s_p &= 2\int\frac{d^3k}{(2\pi)^3}\frac{m^*_N}{E_n}(f^+_n+f^-_n) + \frac{|eB|}{2\pi}\sum_{\nu=0}^{\infty}a_{\nu}\int\frac{dk_z}{2\pi}\frac{m^*_N}{E_{p,\nu}}(f^+_{p,\nu}+f^-_{p,\nu}). 
   \label{scalden}
\end{align}
The equations of motion eqs.(\ref{eom1})-(\ref{eom3}) are to be solved self-consistently for the given values of density, temperature and magnetic field. The baryon chemical potential can also be used as an input parameter in solving them by using the following relation 
\begin{equation}
    \rho_B=\frac{(\mu_B-\mu^*)}{\left(\frac{g_{\omega}^2}{m_{\omega}^2}\right)\left(\frac{\chi^2}{\chi_0^2}\right)}.
    \label{chem1}
\end{equation}
The above relation is obtained by using eq.(\ref{effecpot}) and eq.(\ref{eom1}).

\section{Thermodynamics of the hot magnetized nuclear matter}
\label{sec3}
%The expressions for the total energy density ($\epsilon$) and pressure of the system can be derived from its stress-energy tensor. The energy momentum tensor contains two parts in the presence of a background magnetic field 
%\begin{equation}
 %   T^{\mu\nu}= T^{\mu\nu}_{M} + T^{\mu\nu}_{F}.
 %   \label{t1}
%\end{equation}
%Where, the matter part is written by
%\begin{equation}
%   T^{\mu\nu}_{M} = \epsilon_M u^{\mu}u^{\nu} - P(g^{\mu\nu}-u^{\mu}u^{\nu})+\frac{1}{2}(\mathcal{M}^{\mu\alpha}F^{\nu}_{\alpha}+M^{\nu\alpha}F^{\mu}_{\alpha}) 
 %  \label{t2}
%\end{equation}
%In eq.(\ref{t2}), $\epsilon_M$ denotes the energy density of matter part, $P:$ system's thermodynamic pressure, and $\mathcal{M}^{\mu\nu}$ the magnetization tensor. The field part of the energy momentum tensor has the following contribution from an external magnetic field $B$ along the $\hat{z}$-axis.

The presence of a background magnetic field can have considerable impacts on the thermodynamic properties of nuclear matter. It induces anisotropic structure to the energy-momentum tensor of the system due to the broken rotational symmetry. The pressure splits into the parallel $P^{||}$ and the perpendicular $P^{\perp}$ components along and transverse to the magnetic field direction, respectively, as defined below \cite{pressure1, pressure2}
\begin{equation}
    P^{||}=-\Omega, \quad P^{\perp} = P^{||}-B\mathcal{M} + B^2,
    \label{pressure}
\end{equation}
where, $\mathcal{M}=-\frac{\partial\Omega}{\partial B}$ is the magnetization of the system. The energy density is derived using the thermodynamic relations in the grand canonical ensemble by
\begin{equation}
    \epsilon = \Omega + Ts + \mu_B\rho_B,
    \label{energydensity}
\end{equation}
In eq.(\ref{energydensity}), $s=-\frac{\partial\Omega}{\partial T}$ is the entropy density. The expressions for the entropy density and magnetization can be obtained from the thermodynamic potential as given respectively by 
\begin{multline}
    s = -2\int \frac{d^3k}{(2\pi)^3}\left[ln(1-f_n^+)+ln(1-f_n^-)-\frac{E_n}{T}(f_n^+ + f_n^-)+ \frac{\mu^*}{T}(f_n^+ - f_n^-)\right] \\
    -\frac{|eB|}{2\pi}\sum_{\nu=0}^{\infty}a_{\nu}\int \frac{dk_z}{2\pi}\left[ln(1-f_{p,\nu}^+)+ln(1-f_{p,\nu}^-)-\frac{E_{p,\nu}}{T}(f_{p,\nu}^+ + f_{p,\nu}^-)+ \frac{\mu^*}{T}(f_{p,\nu}^+ - f_{p,\nu}^-)\right],
    \label{entropy}
\end{multline}
and
\begin{multline}
    \mathcal{M}=-B +\frac{e|eB|}{\pi^2}\left[\zeta'(-1,y)+\frac{y^2}{4}+\frac{y}{2}(1-y)lny \right] -\frac{e|eB|}{2\pi^2}y\left[y(1-lny)+ln\Gamma(y)+\frac{1}{2}ln\frac{y}{2\pi}\right]\\ -\frac{eT}{2\pi}\sum_{\nu=0}^{\infty}a_{\nu}\int \frac{dk_z}{2\pi}\left(ln(1-f_{p,\nu}^+)+ln(1-f_{p,\nu}^-)\right) -\frac{e|eB|}{2\pi}\sum_{\nu=0}^{\infty}a_{\nu}\int \frac{dk_z}{2\pi}\frac{\nu}{E_{p,\nu}}(f_{p,\nu}^+ + f_{p,\nu}^-). 
    \label{mag}
\end{multline}
\subsection{Isothermal Compressibility}
In thermodynamics, the concept of isothermal compressibility is a measure of the compression in system's volume due to increasing pressure at a fixed temperature \cite{compressibility1},
\begin{equation}
    K_T=-\frac{1}{V}\left(\frac{\partial V}{\partial P}\right)_T
\end{equation}
The minus sign is to make it a positive quantity. Small values of $K_T$ indicate the matter to be stiffer. The pressure anisotropy at finite magnetic field gives rise to the anisotropic structure in $K_T$ as well. The component parallel to the direction of magnetic field is \cite{compressibility2}
\begin{equation}
    K_T^{||}=-\frac{1}{V}\left(\frac{\partial V}{\partial P^{||}} \right)_T=-\frac{1}{V}\left(\frac{\partial V}{\partial \rho_B} \right)\left(\frac{\partial \rho_B}{\partial P^{||}} \right)_T=\frac{1}{\rho_B^2}\left(\frac{\partial \rho_B}{\partial \mu_B} \right)_T,
    \label{ktpara}
\end{equation}
and the component transverse to the direction of magnetic field is given by
\begin{equation}
    K_T^{\perp} = -\frac{1}{V}\left(\frac{\partial V}{\partial P^{\perp}} \right)_T = \frac{1}{\rho_B(\rho_B-B(\frac{\partial \mathcal{M}}{\partial \mu_B})_T)}\left(\frac{\partial \rho_B}{\partial \mu_B}\right)_T.
    \label{ltperp}
\end{equation}
\subsection{Speed of Sound}
The speed of sound depicts the propagation of disturbances through the medium and depends on the various thermodynamic variables of pressure, energy density, temperature, chemical potential, etc. The speed of sound is, therefore, connected to the equation of state (EOS) of the system and may give significant insights into the space-time evolution of the strongly interacting matter. In general, the squared speed of sound is defined as the change of pressure with respect to the change in energy density for a fixed parameter $\alpha$ \cite{prd105ss},
\begin{equation}
    c_{\alpha}^2=\left (\frac{\partial P}{\partial \epsilon}\right)_{\alpha}
    \label{ss1}
\end{equation}
The in-medium propagation of a compression wave thus needs a specific quantity to be fixed. A variety of choices for $\alpha$ leads to different profiles for the EOS of the nuclear matter system under consideration. In literature, different such cases are considered, such as $\alpha=s/\rho_B, \ s,\ \rho_B,\ T, \ \mu_B$, etc. 
In the relativistic heavy-ion collisions, the fireball created in the early stages of the collisions follows isentropic evolution with fixed entropy density per baryon $\left(\frac{s}{\rho_B}\right)$ through an ideal fluid. %The same conclusion can be drawn in hydrodynamics due to the conservation of energy and baryon number.
Therefore, it is most reasonable to study the speed of sound with the fixed quantity $\alpha=\frac{s}{\rho_B}$ i.e., along the isentropic curve in the present context of hot and dense magnetized symmetric nuclear matter. The squared speed of sound $c^2_{s/\rho_B}$ thus varies with the medium parameters of temperature and baryon chemical potential during the evolution and may give significant insights related to the phase transition, EOS of the medium. % In order to study the intermediate stages of hydrodynamic evolution, the speed of sound with constant entropy density ($c^2_{s}$) or baryon density ($c^2_{\rho_B}$) is considered as the space-time derivative of temperature and chemical potential depend on these quantities. 
In neutron star matter, the quantity $c^2_{T}$ has been studied extensively with considerable experimental importance. 

In the present work, we will study the squared speed of sound along the isentropic curve in a hot magnetized symmetric nuclear matter within the scale-invariance breaking effective model Lagrangian. In the background magnetic field, the speed of sound becomes anisotropic reflecting the nature of pressure anisotropy. It separates into the parallel ($c^{||}_{\alpha}$) and perpendicular ($c^{\perp}_{\alpha}$) directions with respect to the magnetic field. The expression of $c^2_{\alpha}$ in terms of $T$ and $\mu_B$ as obtained from the fundamental thermodynamic relations are given below for the grand canonical ensemble \cite{prc109, prc107ss}:
\begin{multline}
    c^{2}_{\alpha}(T,\mu_B)= c^{2(||)}_{\alpha}(T,\mu_B) = \left[\frac{\partial P^{||}}{\partial \epsilon}\right]_{\alpha}
    = \frac{\left(\frac{\partial P^{||}}{\partial T}\right)_{\mu_B}\left(\frac{\partial \alpha}{\partial \mu_B}\right)_{T}-\left(\frac{\partial P^{||}}{\partial \mu_B}\right)_{T}\left(\frac{\partial \alpha}{\partial T}\right)_{\mu_B}}{\left(\frac{\partial \epsilon}{\partial T}\right)_{\mu_B}\left(\frac{\partial \alpha}{\partial \mu_B}\right)_{T}-\left(\frac{\partial \epsilon}{\partial \mu_B}\right)_{T}\left(\frac{\partial \alpha}{\partial T}\right)_{\mu_B}}, 
    \label{ss2}
\end{multline}
and the component in the perpendicular direction is 
\begin{multline}
    c^{2(\perp)}_{\alpha}(T,\mu_B) = \left[\frac{\partial P^{\perp}}{\partial \epsilon}\right]_{\alpha}= c^{2(||)}_{\alpha}-B\left(\frac{\partial \mathcal{M}}{\partial \epsilon}\right)_{\alpha}
    =c^{2(||)}_{\alpha}-B \frac{\left(\frac{\partial \mathcal{M}}{\partial T}\right)_{\mu_B}\left(\frac{\partial \alpha}{\partial \mu_B}\right)_{T}-\left(\frac{\partial \mathcal{M}}{\partial \mu_B}\right)_{T}\left(\frac{\partial \alpha}{\partial T}\right)_{\mu_B}}{\left(\frac{\partial \epsilon}{\partial T}\right)_{\mu_B}\left(\frac{\partial \alpha}{\partial \mu_B}\right)_{T}-\left(\frac{\partial \epsilon}{\partial \mu_B}\right)_{T}\left(\frac{\partial \alpha}{\partial T}\right)_{\mu_B}}, 
    \label{ss3}
\end{multline}
\\
For the fixed parameters of entropy per baryon density $\alpha=\frac{s}{\rho_B}$ and temperature $\alpha= T$, the above general expressions read
\begin{equation}
c^{2(||)}_{s/\rho_B}=\frac{s\rho_B\left(\frac{\partial s}{\partial \mu_B}\right)_{T}-s^2\left(\frac{\partial \rho_B}{\partial \mu_B}\right)_{T}-\rho_B^2 \left(\frac{\partial s}{\partial T}\right)_{\mu_B} +s\rho_B \left(\frac{\partial \rho_B}{\partial T}\right)_{\mu_B}}{(sT+\mu_B\rho_B)\left[\left(\frac{\partial s}{\partial \mu_B}\right)_{T}\left(\frac{\partial \rho_B}{\partial T}\right)_{\mu_B}- \left(\frac{\partial s}{\partial T}\right)_{\mu_B} \left(\frac{\partial \rho_B}{\partial \mu_B}\right)_{T} \right]},
\label{s1pa}
\end{equation}

\begin{multline}
    c^{2(\perp)}_{s/\rho_B}= \\ c^{2(||)}_{s/\rho_B} - B\frac{\rho_B\left(\left(\frac{\partial \mathcal{M}}{\partial T}\right)_{\mu_B}\left(\frac{\partial s}{\partial \mu_B}\right)_{T}-\left(\frac{\partial \mathcal{M}}{\partial \mu_B}\right)_{T}\left(\frac{\partial s}{\partial T}\right)_{\mu_B}\right)-s \left(\left(\frac{\partial \mathcal{M}}{\partial T}\right)_{\mu_B}\left(\frac{\partial \rho_B}{\partial \mu_B}\right)_{T}-\left(\frac{\partial \mathcal{M}}{\partial \mu_B}\right)_{T}\left(\frac{\partial \rho_B}{\partial T}\right)_{\mu_B}\right)}{(sT+\mu_B\rho_B)\left[\left(\frac{\partial s}{\partial \mu_B}\right)_{T}\left(\frac{\partial \rho_B}{\partial T}\right)_{\mu_B}- \left(\frac{\partial s}{\partial T}\right)_{\mu_B} \left(\frac{\partial \rho_B}{\partial \mu_B}\right)_{T} \right]},
    \label{s1pen}
\end{multline}
and
\begin{equation}
    c^{2(||)}_T=\frac{\rho_B}{T\left(\frac{\partial s}{\partial \mu_B}\right)_{T}+\mu_B \left(\frac{\partial \rho_B}{\partial \mu_B}\right)_{T}},
    \label{stpa}
\end{equation}
\begin{equation}
    c^{2(\perp)}_T=c^{2(||)}_T-B\frac{\left(\frac{\partial \mathcal{M}}{\partial \mu_B}\right)_{T}}{T\left(\frac{\partial s}{\partial \mu_B}\right)_{T}+\mu_B \left(\frac{\partial \rho_B}{\partial \mu_B}\right)_{T}}.
    \label{stpen}
\end{equation}

 \section{Results and Discussions}
 \label{sec4}
In this section, the in-medium behaviors of several thermodynamic quantities are presented for different values of temperature ($T$), magnetic field ($|eB|$) and baryon chemical potential ($\mu_B$) in a hot magnetized symmetric nuclear matter. 
The field equations are solved self-consistently for the mean-meson fields $\sigma$, $\chi/\chi_0$ at the given values of $\mu_B$, $T$ and $|eB|$. %In the context of non-central heavy ion collisions, strong magnetic fields are estimated to be produced along the perpendicular direction to the reaction plane. The magnitude of the produced magnetic field in the center-of-mass frame depends on various collision parameters of impact parameter, center-of-mass energy per nucleon, electric charge of the colliding nuclei, etc.
\textcolor{black}{ The estimated field strength in the ultra-relativistic non-central Au-Au collisions at RHIC can be of the order of $\sim10^{18}-10^{19}$ G $\approx 0.02-0.2\ \text{GeV}^2$, and $\sim10^{20}$ G for the Pb-Pb collisions at the LHC \cite{sokov, deng}. %The magnetic field is produced at the initial stages of the collisions which eventually decay very rapidly within a first few fm/c. The decay may sufficiently delayed due to an expected finite conductivity of the medium such that a finite field strength may still persist in the subsequent hadronic phases. 
The magnetic field in the interior of the neutron star matter can go up to $\sim 10^{18}$ G \cite{shapiro}. %, whereas the surface magnetic fields of the magnetars are only of the order of $10^{14}-10^{15}$ Gauss \cite{thompson}.
In the CBM experiment at FAIR accelerators SIS100 and SIS300, nuclear matter is expected to be produced at high baryon densities $\rho_B \approx 5\rho_0-10\rho_0$ corresponding to the beam energy of $5A-20A$ GeV \cite{cbm}. The density of the neutron star matter varies with its different layers. For hybrid stars, there may be an external layer of ordinary nuclear matter up to a density of $\sim 2\rho_0$, whereas the density at the internal core can go beyond that limit \cite{pressure2}. % The density at the core of neutron star can reach more than $2\rho_0$ with an external layer of ordinary nuclear matter up to a density of $2\rho_0$ . %According to several effective model studies, the critical temperature for the chiral transition i.e., a QCD phase transition from the hadronic to quark matter regime lies in the 
In the LQCD studies at zero baryon density, the crossover transition between the hadron gas and QGP is found to be around $T_c\approx 156.5\pm1.5$ MeV, which corresponds to a minimum in the value of sound speed \cite{tc}. Hence, in the study of the thermodynamic properties of magnetized hot and dense nuclear matter, baryon density is kept within a range of $\rho_B =0-6\rho_0$ corresponding to $\mu_B\approx 700-1500$ MeV, for temperatures up to $T=100$ MeV and constant magnetic field strength up to $0.08 \ \text{GeV}^2 \sim 4\times 10^{18}$ G.} 

In the present study, we have considered summation over 500 Landau energy levels of protons to find the solutions for the meson fields at finite values of $\mu_B$, $T$ and $|eB|$ by maintaining the convergence of the results. \textcolor{black}{The number of Landau levels (L) required to find the stable solutions for the effective nucleon mass $m^*_N$ and other thermodynamics variables vary slightly with the medium parameters of $\mu_B$, $T$ and $|eB|$. The thermodynamic quantities do not change with increasing Landau levels from $L=500$ to $505,\ 510, $ up to $600$ at higher values of temperature and chemical potential, for e.g., at $T=100$ MeV and $\mu_B=1200$ MeV ($\rho_B \approx 3.7\rho_0$)}.% For $|eB|=0.04\ GeV^2$ and $\rho_B=3.5\rho_0$, the number of required Landau levels is changed from $L=4$ to $54$ as the temperature changes from $T=10$ to $100$ MeV. At a lower baryon density $\rho_B=\rho_0$, the corresponding shift is obtained from $L=3$ to $L=65$. At a lower magnetic field strength $|eB|=0.02\ GeV^2$ and high baryon density $\rho_B=6\rho_0$, the number of filled Landau levels $L=9$ at $T=10$ MeV changes to $L=110$ at $T=100$ MeV. This can be compared to the low density case at $\rho_B=2\rho_0$, where a shift of $L=8-125$ is obtained corresponding to the shift in temperature from $T=10-100$ MeV. Thus, the number of Landau levels needed to find the convergence remain well below the considered value of $L=500$ for the range of baryon density and temperature studied here. Thus, the representative value $L=500$ is chosen for convenience as the thermodynamic quantities do not change with $L$ after the convergence limit is reached as mentioned above for different set of low and high temperature and chemical potential. Although, for strong magnetic field the contribution of the lowest Landau level is only effective. At weak fields, the effects of higher Landau levels need to be considered as well.

The effective mass of the nucleon $m^*_N$ is thus studied along with the scaled magnetization ($\mathcal{M}_{scaled}$), anisotropic pressure ($P$), isothermal compressibility ($K_T$) and the squared speed of sound ($c^2_{s/\rho_B}$) in the thermomagnetic symmetric nuclear matter system under consideration. 

\subsection{Effective Mass of the Nucleon}
\begin{figure}[h!]%
    \centering
    \subfloat[\centering ]{{\includegraphics[width=8.4cm]{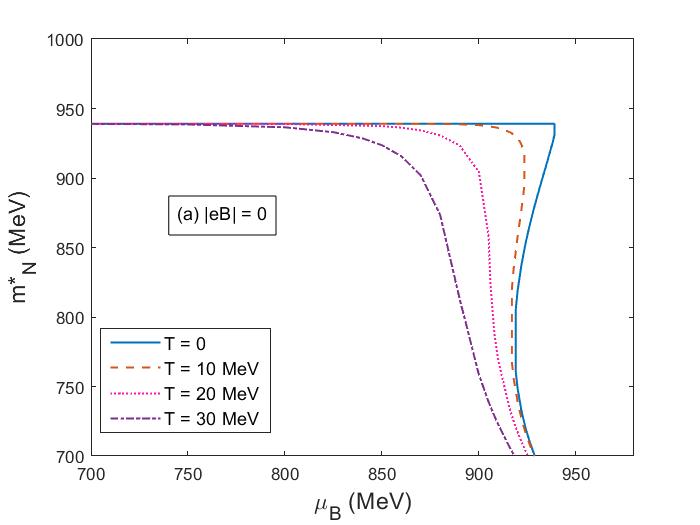} }}%
    \hspace{-0.8cm}
    \subfloat[\centering ]{{\includegraphics[width=8.4cm]{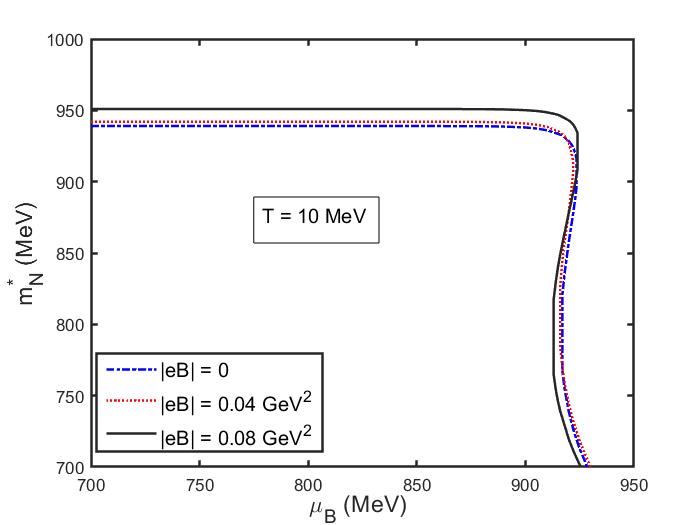} }}%
    \vspace{-0.4cm}
    \caption{\raggedright{ The effective mass of the nucleon $m^*_N$ (in MeV) is plotted as a function of the baryon chemical potential $\mu_B$ (in MeV) for (a) different temperatures $T=0,10,20,30$ MeV at zero magnetic field $|eB|=0$ and (b) different magnetic field strengths $|eB|=0,0.04,0.08\ \text{GeV}^2$ at a fixed temperature $T=10$ MeV.}}%
    \label{m1}%
\end{figure}

The effective mass of the nucleon $m^*_N$ is calculated from eq.(\ref{effecmass}) by using the solutions for the mean meson fields $\sigma$ and $\chi/\chi_0$ at the given values of $\mu_B$, $T$, and $|eB|$. The vacuum nucleon mass is taken to be $m_N=939$ MeV in the calculation. The effects of temperature are incorporated through the fermion (antifermion) distribution functions of protons and neutrons. The impacts of magnetic field are considered through the Landau energy levels of protons along with the magnetized vacuum contribution of nucleons in the symmetric nuclear matter environment. \textcolor{black}{In the present study, the solutions are found for both zero and finite temperatures $(T=0,10,20,30)$ at $|eB|=0, 0.04, 0.08\ \text{GeV}^2 \ (0.0195\ \text{GeV}^2\approx 10^{18}$ G) of magnetic field strength, in a range of baryon chemical potential.} In fig.\ref{m1}, the effective nucleon mass $m^*_N$ (in MeV) is plotted with the variation in baryon chemical potential $\mu_B$ (in MeV) at (a) $|eB|=0$, for $T=0,10,20,30$ MeV, and (b) $T=10$ MeV for $|eB|\approx0, 2\times 10^{18}, 4\times 10^{18}$ G. In the vacuum, the baryon number density is zero i.e., $\rho_B=0$, equivalently $\omega_0=0$ from eq.(\ref{eom1}) since the ratio $\frac{\chi}{\chi_0}$ should be approximately $1$ in vacuum. The change in the scalar glueball field obtained in terms of the ratio $\chi/\chi_0$ is marginal with respect to the change in $\mu_B$. There are some general features of the solutions at both $T=0$ and $T\neq0$ for the considered magnetic field strengths. Let us first understand the horizontal segment of the plots. If the effective chemical potential ($\mu^*$) (corresponds to the effective energy $E^*_N=\sqrt{m^{*2}_N+k^2}$ at $T=0$, $|eB|=0$) is less than the effective mass ($m^*_N$) then there must not be any medium contribution to the solutions \cite{haber}.
\begin{figure}[h!]%
    \centering
    \subfloat[\centering ]{{\includegraphics[width=8.4cm]{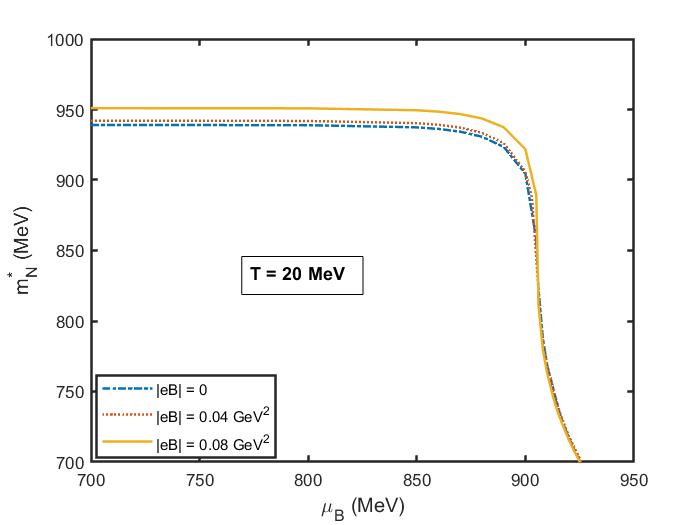} }}%
    \hspace{-0.8cm}
    \subfloat[\centering ]{{\includegraphics[width=8.4cm]{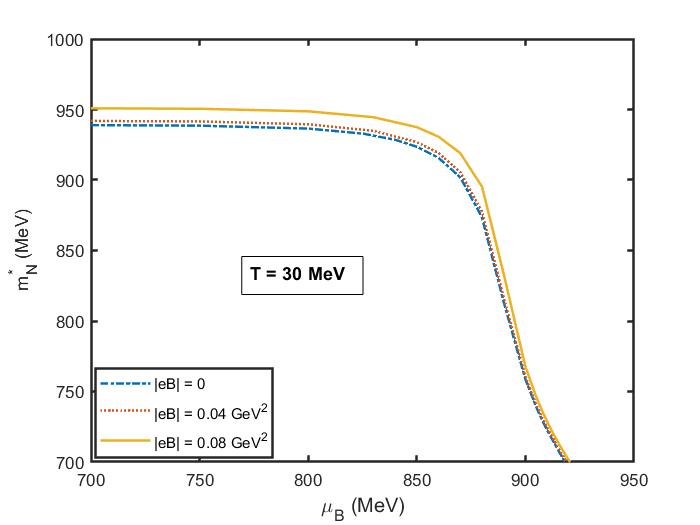} }}%
    \vspace{-0.4cm}
    \caption{\raggedright{ The effective mass of the nucleon $m^*_N$ (in MeV) as a function of the baryon chemical potential $\mu_B$ (in MeV) for the field strengths of $|eB|=0,0.04,0.08\ \text{GeV}^2$ at the temperature (a) $T=20$ MeV, and (b) $T=30$ MeV.}}%
    \label{m2}%
\end{figure}
 This is similar to the solutions obtained in vacuum ($\rho_B=0$) case, where $\mu^*=\mu_B$ from eq.(\ref{effecpot}). Hence, the effective mass of the nucleon does not vary with $\mu_B$ for $\mu_B<m^*_N$ for vacuum solutions, giving rise to a horizontal line segment as in figs.(\ref{m1})-(\ref{m2}). \textcolor{black}{However, the monotonic rise of the vacuum solution with magnetic field depicts the well-known effect of magnetic catalysis \cite{prd98}, observed in fig.(\ref{m1}b) and figs.(\ref{m2}a-\ref{m2}b) for the increasing field strength $|eB|=0, 0.04$ and $0.08 \ \text{GeV}^2$ at a fixed temperature $T=10,20,30$ MeV.} The medium terms contribute when the effective energy is larger than the effective mass and $\rho_B\approx \omega\neq 0$ in eq.(\ref{eom1}). 
In figs.(\ref{m1}), at $T=0$ and $T=10$ MeV, there exist three solutions for the nucleon mass for a particular region of $x$-axis or $\mu_B$ values. A first-order phase transition from vacuum to nuclear matter may occur in this region, within the context of present model. The onset of nuclear matter depends on the relative free energies between the medium and vacuum as it has been illustrated in ref.\cite{haber}. Similar trend was obtained for the Walecka model studies at zero as well as finite temperature nuclear matter \cite{haber, prc109}. \textcolor{black}{Thus, the effective nucleonic mass $m^*_N$ increases with magnetic field at a fixed temperature within a range of baryon chemical potential $\mu_B=700-1200$ MeV in fig.(\ref{m1}b) and figs.(\ref{m2}a)-(\ref{m2}b). However, the variation of $m^*_N$ as a function of $\mu_B$ for zero magnetic field is shown for different temperatures in fig.(\ref{m1}a). This indicates the nature of the vacuum to nuclear matter phase transition region at low temperatures, suitable for the nuclear matter study within the current effective model approach.} 
\begin{figure}[h!]%
    \centering
    \subfloat[\centering ]{{\includegraphics[width=8.4cm]{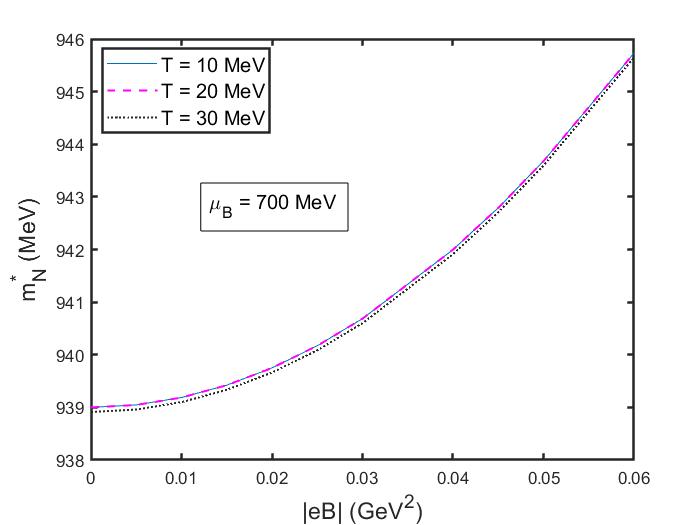} }}%
    \hspace{-0.8cm}
    %\subfloat[\centering ]{{\includegraphics[width=8.4cm]{mc9.jpg} }}%
   %  \hspace{-0.8cm}
    \subfloat[\centering ]{{\includegraphics[width=8.4cm]{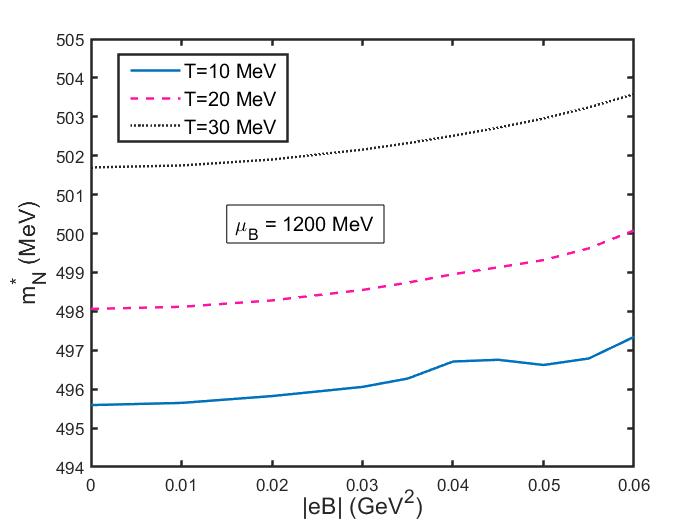} }}%
    \vspace{-0.4cm}
    \caption{\raggedright{ The effective nucleon mass $m^*_N$ (in MeV) is plotted as a function of the external magnetic field $|eB|$ (in $\text{GeV}^2$) for $T=10,20,30$ MeV temperatures at (a) $\mu_B=700$ MeV, and (b) $\mu_B=1200$ MeV.}}%
    \label{m3}%
\end{figure}
In fig.(\ref{m3}), the effective nucleon mass is plotted as a function of magnetic field for three different temperatures $T=10,20,30$ MeV at (a) $\mu_B=700$ MeV, and (b) $\mu_B=1200$ MeV. In fig.(\ref{m3}a), the mass increases monotonically with magnetic field at lower chemical potential, giving rise to the well-known phenomenon of magnetic catalysis. At higher chemical potential, considerable change of $(3-4)$ MeV in $m^*_N$ is obtained with respect to the change in temperature as in fig.(\ref{m3}b), in comparison to the low chemical potential behavior where there is barely any change with temperature.

 \textcolor{black}{In solving the self-consistent equations for the scalar meson fields $\sigma$ and $\chi/\chi_0$, the nuclear matter density or baryon density $\rho_B$ in eq.(\ref{noden}) depends on the given input parameters $\mu_B$, $T$ and $|eB|$. Now, for a fixed value of $\mu_B=1200$ MeV, as $T$ changes from $10$ MeV to $30$ MeV for a particular value of $|eB|=0.01\ \text{GeV}^2$, $\rho_B$ changes slightly from $3.68\rho_0$ to $3.72\rho_0$. For $T=10$ MeV and $\mu_B=1200$ MeV, it remains almost unchanged from $3.68\rho_0-3.69\rho_0$ as the field strength changes from $|eB|=0.02-0.05\ \text{GeV}^2$. In fig.(\ref{m3}a), the chosen $\mu_B=700$ MeV corresponds to the vacuum case as it is also observed from the horizontal segment in figs.(\ref{m1}-\ref{m2}), for the given range of temperature as well as magnetic field strengths. Thus, in figs.(\ref{m3}a)-(\ref{m3}b) the effects of magnetic catalysis is shown for two extreme values of nuclear matter densities $\rho_B=0$ $(\sim \mu_B=700$ MeV) and $ 4\rho_0\sim$ ($\mu_B=1200$ MeV) for a set of temperature and magnetic field appropriate for the dense nuclear matter study within the present approach.}
 
\subsection{Anisotropic Pressure}
\begin{figure}[h!]%
    \centering
    \subfloat[\centering ]{{\includegraphics[width=8.4cm]{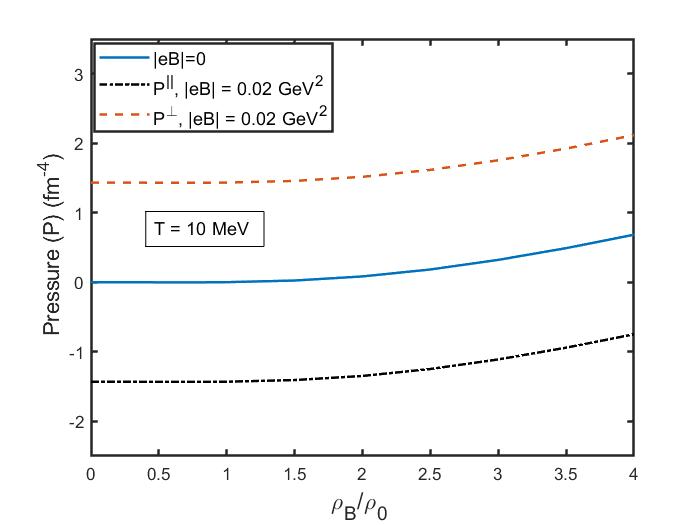} }}%
    \hspace{-0.8cm}
    \subfloat[\centering ]{{\includegraphics[width=8.4cm]{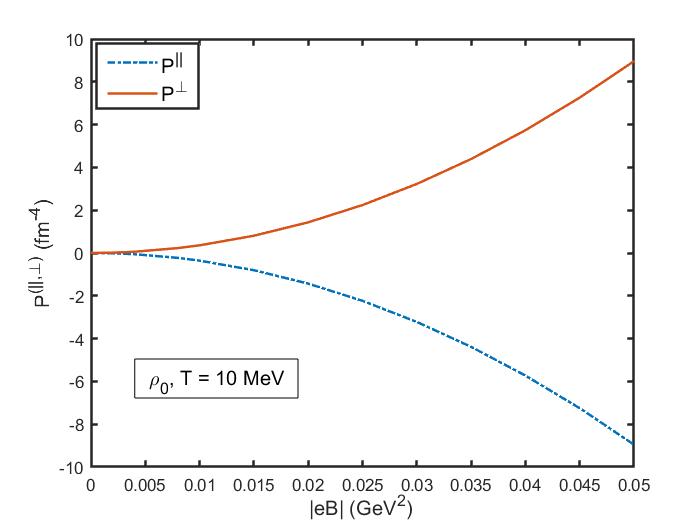} }}%
    \vspace{-0.4cm}
    \caption{\raggedright{The pressure parallel and perpendicular to the field direction $P^{(||,\perp)}$ (in $\text{fm}^{-4}$) are plotted as functions of (a) baryon density $\rho_B$ (in units of $\rho_0$) and (b) magnetic field (in $\text{Ge}V^2$) for $T=10$ MeV. In plot (a), magnetic field is fixed at $|eB|=0.02\ \text{GeV}^2$, and the zero-field case is shown for comparison (solid line). In plot (b), the splitting with respect to the field direction are shown for a fixed density $\rho_B=\rho_0$. }}%
    \label{p1}%
\end{figure}

The anisotropic pressure is plotted as a function of baryon density and magnetic field in fig.(\ref{p1}a) and fig.(\ref{p1}b), respectively at $T=10$ MeV. The behavior is shown for pressures parallel and perpendicular to the field direction, denoted by $P^{||}$ and $P^{\perp}$, respectively and plotted in units of $\text{fm}^{-4}$. In fig.(\ref{p1}a), the zero-field case is also shown for comparison. For finite magnetic field, the splitting between these components are shown at the nuclear matter saturation density $\rho_0$ for $T=10$ MeV. There is no splitting for $|eB|=0$ due to the absence of any field contribution, as it is obvious from eq.(\ref{pressure}). The magnitude of the parallel pressure $P^{||}$ comes out to be negative, which indicates an instability arising due to the negative contribution of the field energy density to the pressure along the field direction \cite{monika, prd101}. The amount of splitting increases with magnetic field strength as shown in fig.\ref{p1}(b). 

\subsection{Magnetization}
\begin{figure}[h!]%
    \centering
    \subfloat[\centering ]{{\includegraphics[width=8.4cm]{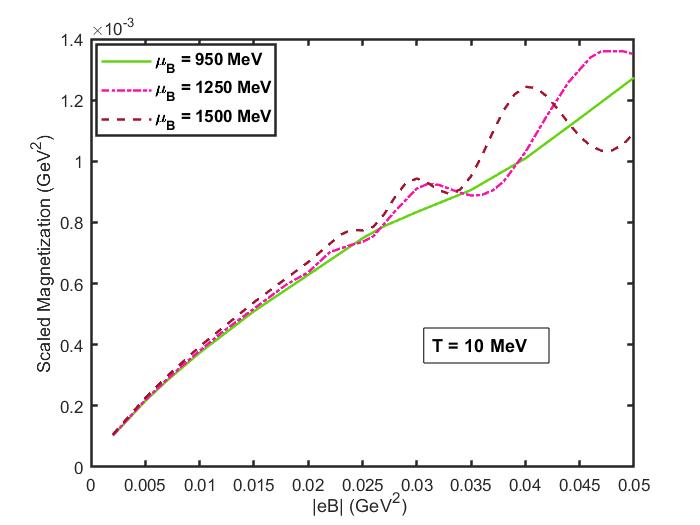} }}%
    \hspace{-0.8cm}
    \subfloat[\centering ]{{\includegraphics[width=8.4cm]{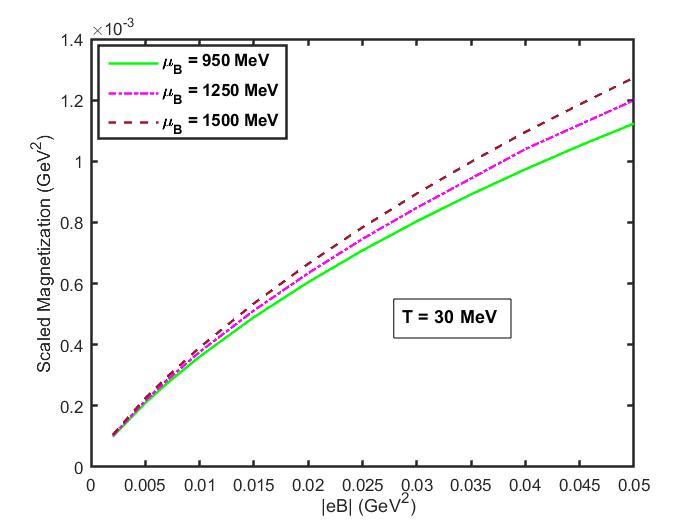} }}%
    \vspace{-0.4cm}
    \caption{\raggedright{The scaled magnetization $\mathcal{M}_{scaled}$ (in $ \text{Ge}V^2$) is plotted as a function of $|eB|$ (in $ \text{GeV}^2$) for $\mu_B=950,1250, 1500$ MeV at (a) $T=10$ MeV, and (b) $T=30$ MeV.}}%
    \label{mag1}%
\end{figure}

In the presence of a background magnetic field, the nuclear matter system becomes magnetized and leads to a finite value of magnetization. It is defined by the change in thermodynamic potential of the system with respect to a change in magnetic field as given by eq.(\ref{mag}). For convenience, we define the scaled magnetization as  
\begin{equation}
    e\mathcal{M}_{scaled} = \mathcal{M}+B
    \label{scalemag}
\end{equation}
\textcolor{black}{The current study is mainly concerned with the nuclear matter properties below the baryonic density of $\rho_B=8\rho_0$ at which nucleon core overlaps. In fig.\ref{mag1}, the scaled magnetization $\mathcal{M}_{scaled}$ (in $\text{GeV}^2$) is plotted as a function of the magnetic field $|eB|$ (in $\text{GeV}^2$) for $\mu_B=950,1250,1500$ MeV at (a) $T=10$ MeV, and (b) $30$ MeV. The baryon densities corresponding to $\mu_B=950,1250,1500$ MeV are approximately $\rho_B\approx 1.6\rho_0, 4\rho_0,6\rho_0$, respectively, which are less than the limiting value of $8\rho_0$. In fig.(\ref{mag1}a), the positive valued scaled magnetization show oscillatory pattern at large magnetic field strengths for higher chemical potentials. This nature originates from the Landau energy levels of the electrically charged protons due to their orbital motion in a background magnetic field. However, the effect of Landau levels is expected to be washed out at high temperature and chemical potentials in particular at high temperature when the temperature effect dominates over the field contribution.} 

%The number of Landau levels (L) required to find stable solutions for the nucleon mass $m^*_N$ along with the thermodynamics variables of $P^{||,\perp}$, $\mathcal{M}_{scaled}$, $K_T^{||,\perp}$, $C_{s/\rho_B}^2$ are varied slightly with the medium parameters of $\mu_B$, $T$ and $|eB|$. Although the magnitudes of these quantities do not change as we increase the Landau levels from $\nu=500$ to $505,\ 510, $ up to $600$ for higher values of temperature and chemical potentials for e.g., at $T=100$ MeV and $\mu_B=1200$ MeV corresponds to $\rho_B = \rho_0$. At $T=10$ MeV and $\rho_B=\rho_0$, For $|eB|=0.04\ GeV^2$, the number of Landau levels needed to find the convergence is changed from $L=4$ for $T=10$ MeV to $L=54$ for $T=100$ MeV at a high baryon density of $\rho_B=3.5\rho_0$. Similarly, at a low baryon density $\rho_0$, this shift is obtained from $L=3$ at $T=10$ MeV to $L=65$ at $T=100$ MeV. At a lower magnetic field strength $|eB|=0.02\ GeV^2$, the no. of filled Landau levels leading to the convergence $L=9$  at $T=10$ MeV changes to $L=110$ for $T=100$ MeV at a high baryon density $\rho_B=6\rho_0$.1.

In the current hadronic model approach, this particular nature of oscillation ceases with increasing temperature, as shown in fig.(\ref{mag1}b) at $T=30$ MeV. In the present analysis, the field-dependent vacuum contribution does not impose any renormalization/regularization artifact on the thermodynamic quantities of the system. Hence, the oscillatory behaviors obtained in the magnetization are independent of any such renormalization scheme.%, which might be the case in many non-renormalizable NJL type models analysis. 
\subsection{Isothermal Compressibility}
%We have investigated the isothermal compressibility in a hot magnetized isospin symmetric nuclear matter within a scale-symmetry breaking effective model Lagrangian. 
The characteristic nature of variation of the isothermal compressibility with chemical potential is considered to be a potential indicator for the stiffness of the equation of state (EOS). Thus, analysis of this property in a magnetized nuclear matter at very low temperatures can have important significance in the context of many compact  astrophysical objects, e.g., neutron star, magnetars etc.    
\begin{figure}[h!]%
    \centering
    \subfloat[\centering ]{{\includegraphics[width=8.4cm]{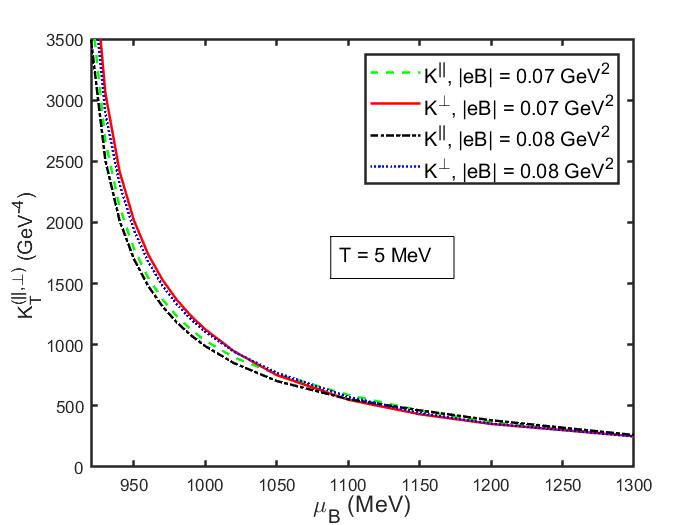} }}%
    \hspace{-0.8cm}
    \subfloat[\centering ]{{\includegraphics[width=8.4cm]{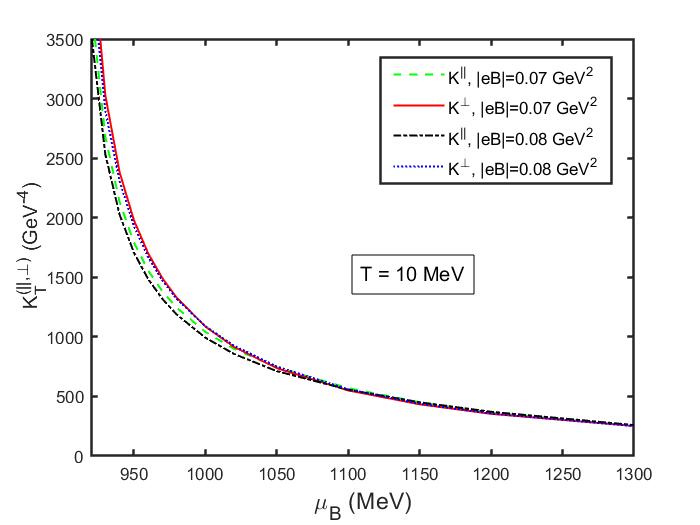} }}%
    \vspace{-0.4cm}
    \caption{\raggedright{The parallel and perpendicular components of isothermal compressibility $K_T^{(||,\perp)}$ (in $ \text{GeV}^{-4}$) are plotted as functions of $\mu_B$ (in MeV) for $|eB|=0.07,0.08\ \text{GeV}^2$, at (a) $T=5$ MeV, and (b) $T=10$ MeV of temperatures.}}%
    \label{kt}%
\end{figure}

In the presence of an external magnetic field, the isothermal compressibility $K_T$ separates into two components, namely in the parallel and perpendicular directions with respect to the magnetic field, denoted by $K_T^{||}$ and $K_T^{\perp}$, respectively. Relative to the case of isotropic compressibility at zero magnetic field, the present study shows anisotropy at finite magnetic field for different values of $\mu_B$ and $T$. These components $K_T^{(||,\perp)}$ (in $\text{GeV}^{-4}$) are plotted in figs.(\ref{kt}a) and (\ref{kt}b) as functions of the baryon chemical potential $\mu_B$ (in MeV) at $T=5,10$ MeV for $|eB|=0.07,\ 0.08\ \text{GeV}^2$. It is observed that, the compressibility is different in the parallel and perpendicular directions with respect to the magnetic field. The magnitudes of both the components decrease with increasing chemical potential. The value reduces to a large extent at higher chemical potentials, indicating a much stiffer equation of state in this regime. A considerable dependence on the magnetic field strength is observed at lower values of $\mu_B$. While at higher $\mu_B$ values, this dependence becomes much weaker. In this region, presumably the number density is so high that the stiffness of the equation of state cannot change further. However, the component in the parallel direction decreases by a larger amount as compared to the perpendicular direction, leading to a stiffer equation of state parallel to the magnetic field direction. Similar behaviors are obtained at different values of temperature and magnetic field strengths considered here. The pressure anisotropy thus reflects into the in-medium behavior of compressibility by giving rise to two components with respect to the direction of magnetic field. 
\begin{figure}[h!]
    \centering
    \includegraphics[width=0.6\linewidth]{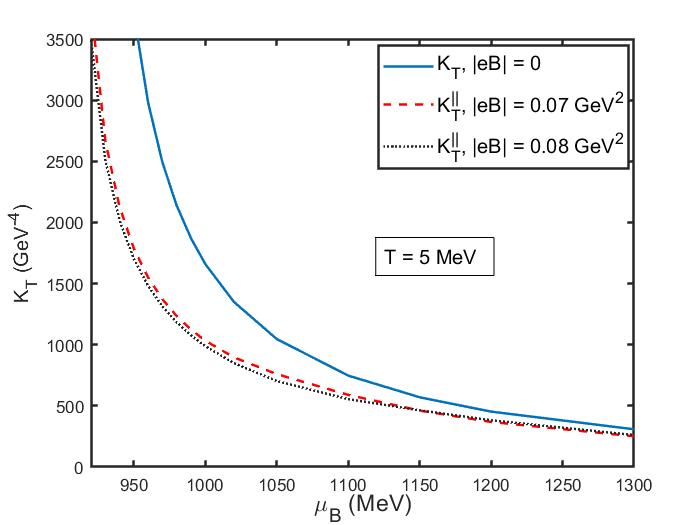}
    \caption{\raggedright{ Parallel component of the compressibility is plotted as a function of $\mu_B$ at $T=5$ for finite magnetic field strengths $|eB|=0.07,0.08\ \text{GeV}^2$ at $T=5$ MeV. For comparison, the compressibility at zero magnetic field is shown by solid line.}}
    \label{k2}
\end{figure}
In fig.(\ref{k2}), the isothermal compressibility $K_T$ (its $K_T^{||}$ component) is plotted as a function of $\mu_B$ at zero (finite) magnetic field for a particular temperature of $T=5$ MeV. The variation of $K_T$ with $\mu_B$ at zero magnetic field has similar trend as the finite magnetic field case except for a larger magnitude of compressibility. It is, therefore, observed that to some extent magnetic field leads to a much stiffer equation of state as compared to the case of zero magnetic field.  

\textcolor{black}{We consider two representative values of the field strength $|eB|=0.07,$ and $0.08\ \text{GeV}^2$ $(\sim10^{18}\ \text{G}$ ) in the study of isothermal compressibility. This particular choice is convenient for the study of neutron star matter interior, where the field strength can lie in the range of  $\sim(10^{18}-10^{20}\ \text{G})$, and it can be of the order of $\sim(10^{14}$  -$10^{15}\ \text{G}$ ) at the surface layers of some magnetars. The components of $K_T^{||,\perp}$ with respect to the direction of magnetic field can be compared for any of these chosen field strengths. However, the magnitude of isothermal compressibility is compared for different field strengths with respect to the zero field case in fig.(\ref{k2}) and is observed to decrease slightly with increasing magnetic field, indicating a stiffer system of nuclear matter. %Strong magnetic field affects the EOS and pressure within the body of the star. % It primarily leads to expansion and a decrease in average density and compactness of the star rather than making it more resistant to compression.
The presence of magnetic field affects the surface geometry and hence the volume of the matter. In fact a high magnetic field makes the body less compressible, particularly in the outer layers of neutron star. The effect of lowering the compressibility is particularly noticeable in the crust of the neutron star system where magnetic field can significantly influence the properties of nuclear matter (EOS, pressure, energy density etc.). The compressibility measures a substance's ability to change in volume under pressure, a reverse effect of incompressibility. The concept of incompressibility is related to the resistance to compression under a change in volume by applying pressure. Therefore, low compressibility implies that the matter is highly incompressible i.e., the matter becomes more resistant with respect to a volume change.}

\subsection{Speed of Sound}
\begin{figure}[h!]%
    \centering
    \subfloat[\centering ]{{\includegraphics[width=8.4cm]{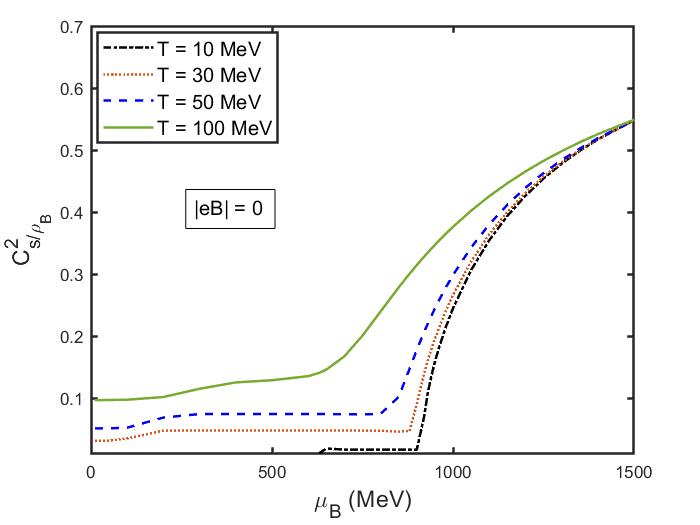} }}%
    \hspace{-0.8cm}
    \subfloat[\centering ]{{\includegraphics[width=8.4cm]{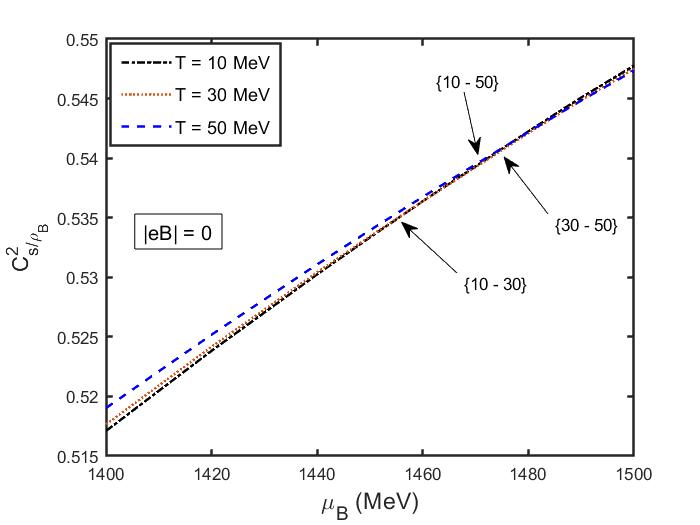} }}%
    \vspace{-0.4cm}
    \caption{\raggedright{Squared speed of sound $c^2_{s/\rho_B}$ along the isentropic curve as a function of baryon chemical potential for various fixed temperature $T=10,30,50,100$ MeV at zero magnetic field. Plot (b) is a zoomed in version of plot (a) for $T=10,30,50$ MeV in the $(1400-1500)$ MeV range of $\mu_B$-axis.}}%
    \label{c1}%
\end{figure}

In this subsection, the squared speed of sound is studied along the isentropic curve for different values of $\mu_B(\rho_B)$, $T$ and $|eB|$ in a symmetric nuclear matter. The basic thermodynamic quantities in eqs.(\ref{s1pa})-(\ref{s1pen}) required in the calculation of $c^{2(||,\perp)}_{s/\rho_B}$ are derived from the thermodynamic potential in eq.(\ref{thermopot}). In fig.(\ref{c1}a), $c^2_{s/\rho_B}$ is plotted as a function of $\mu_B$ for $T=10,30,50,100$ MeV at zero magnetic field. There is no anisotropic structure observed in the absence of magnetic field. 

\begin{figure}[h!]%
    \centering
    \subfloat[\centering ]{{\includegraphics[width=8.4cm]{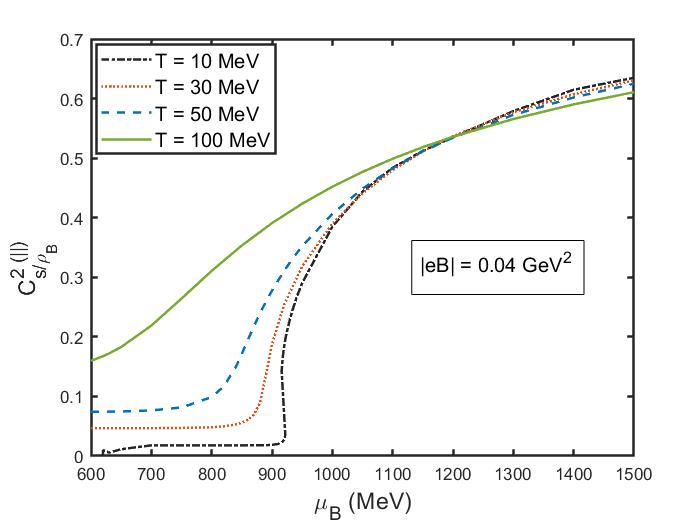} }}%
    \hspace{-0.8cm}
    \subfloat[\centering ]{{\includegraphics[width=8.4cm]{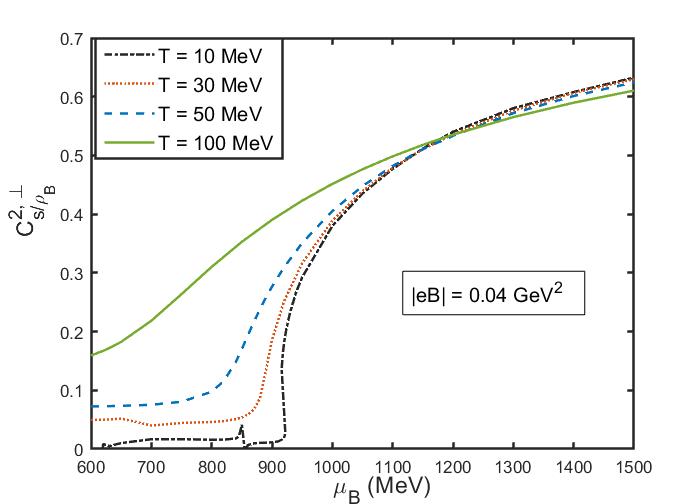} }}%
    \vspace{-0.4cm}
    \caption{\raggedright{(a) The parallel $(c^{2(||)}_{s/\rho_B})$ and (b) perpendicular $(c^{2(\perp)}_{s/\rho_B})$ components of the squared sound speed are plotted as functions of $\mu_B$ (in MeV) for $T=10,30,50,100$ MeV at $|eB|=0.04\ \text{GeV}^2$. }}%
    \label{c2}%
\end{figure}

%\begin{figure}[h!]
 %   \centering
 %   \includegraphics[width=0.6\linewidth]{c0.jpg}
  %  \caption{\raggedright{Squared speed of sound $C^2_{s/\rho_B}$ along the isentropic curve as a function of baryon chemical potential for various fixed temperature $T=10,30,50,100$ MeV in the absence of magnetic field.}}
  %  \label{c1}
%\end{figure}
At $T=10$ MeV its value suddenly starts rising around $\mu_B=920$ MeV, reflecting the nature of first-order phase transition as depicted in fig.(\ref{m1}a) for $|eB|=0$. This point is shifted towards lower chemical potentials with increasing temperature. The magnitude of $c^2_{s/\rho_B}$ increases with temperature in the lower region of $\mu_B$ values, which shows opposite trend at higher values of chemical potentials. This may be due to the suppression of the density part over the temperature one in the denominator of eq.(\ref{s1pa}). The point of transition does not correspond to a fixed $\mu_B$, rather it changes with temperature. In fig.(\ref{c1}b), the transition region is shown for the set of temperatures $T=10,30,50$ MeV, which denote the crossing between the lines correspond to these temperatures. It can be inferred that this shift occurs at a higher chemical potential with increasing temperature for e.g., the transition point for a temperature change of $10$ MeV to $30$ MeV (denoted by $\{10-30\}$ in fig.(\ref{c1}b)) is located at a lower $\mu_B$ value than the corresponding shift between $10$ MeV to $50$ MeV temperature. The plots in figs.(\ref{c2}a)-(\ref{c2}b) illustrate the variation of the (a) parallel ($c^{2(||)}_{s/\rho_B}$) and (b) perpendicular ($c^{2(\perp)}_{s/\rho_B}$) squared sound speed as functions of $\mu_B$ for a set of temperature similar to fig.\ref{c1}, at finite magnetic field strength of $|eB|=0.04\ \text{GeV}^2$. The overall magnitude increases in comparison to the zero field value of sound speed. Similar behaviors are obtained with respect to the change in temperature. However, the point at which the squared sound speed starts decreasing with temperature, shifts towards lower chemical potential relative to the corresponding transition point at zero magnetic field. The components of $c^{2}_{s/\rho_B}$ in the parallel and perpendicular directions with respect to the magnetic field show similar behavior with the change in baryon chemical potential and temperature. A slight deviation is observed at low temperatures $T=10,30$ MeV towards lower chemical potential values in the perpendicular direction relative to the parallel one.

%The physical effects of crossings can be understood better if we analyze the variation of $c_{s/\rho_B}^2$ with respect to the change in temperature as plotted in figs.(\ref{c3}-\ref{c5}). 
\begin{figure}[h!]%
    \centering
    \subfloat[\centering ]{{\includegraphics[width=8.4cm]{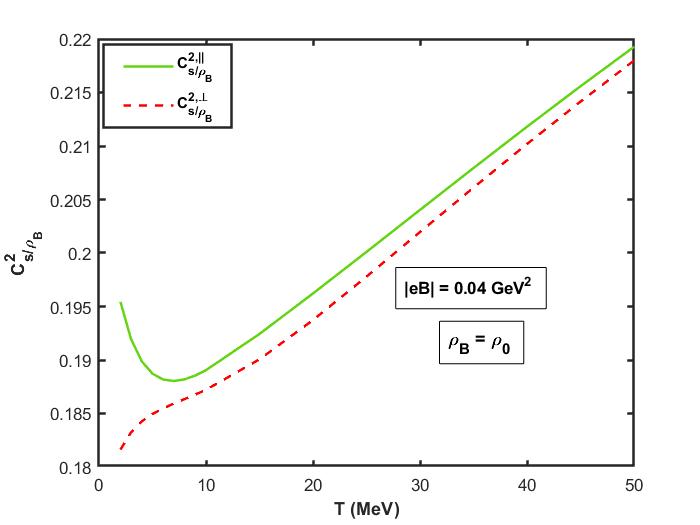} }}%
    \hspace{-0.8cm}
    \subfloat[\centering ]{{\includegraphics[width=8.4cm]{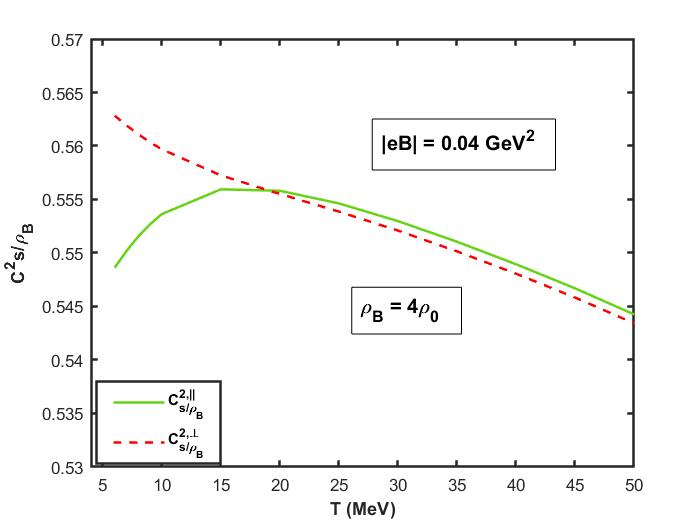} }}%
    \vspace{-0.4cm}
    \caption{\raggedright{Squared speed of sound $c^2_{s/\rho_B}$ as a function of temperature $T$ (MeV) at (a) $\rho_B=\rho_0$ and (b) $\rho_B=4\rho_0$ for $|eB|=0.04\ \text{GeV}^2$.}}%
    \label{c3}%
\end{figure}
%\begin{figure}[h!]%
 %   \centering
 %   \subfloat[\centering ]{{\includegraphics[width=8.4cm]{fig20.jpg} }}%
 %   \hspace{-0.8cm}
  %  \subfloat[\centering ]{{\includegraphics[width=8.4cm]{fig21.jpg} }}%
  %  \vspace{-0.4cm}
 %   \caption{\raggedright{Squared speed of sound $c^2_{s/\rho_B}$ as a function of temperature $T$ (MeV) at (a) $\rho_B=3\rho_0$ and (b) $\rho_B=4\rho_0$ for $|eB|=0.04GeV^2$.}}%
  %  \label{c4}%
%\end{figure}
%\begin{figure}[h!]%
 %   \centering
 %   \subfloat[\centering ]{{\includegraphics[width=8.4cm]{fig22.jpg} }}%
  %  \hspace{-0.8cm}
  %  \subfloat[\centering ]{{\includegraphics[width=8.4cm]{fig23.jpg} }}%
  %  \vspace{-0.4cm}
  %  \caption{\raggedright{Squared speed of sound $c^2_{s/\rho_B}$ as a function of temperature $T$ (MeV) at (a) $\rho_B=5\rho_0$ and (b) $\rho_B=6\rho_0$ for $|eB|=0.04GeV^2$.}}%
%    \label{c5}%
%\end{figure}

\textcolor{black}{It is shown in fig.(\ref{c1}a) that $c^2_{s/\rho_B}$ increases steadily with increasing $\mu_B$ for a particular value of temperature after the region of sudden rise in $c^2_{s/\rho_B}$ for zero magnetic field. For smaller $\mu_B$ values, squared speed of sound increases with increasing temperature from $T=10-100$ MeV as the pressure and energy density are dominated by the temperature driven effects. However, in the high $\mu_B$ regime, its magnitude slightly decreases with rising temperature. The temperature driven influence on the pressure and energy density is shifted towards the density driven influence on these thermodynamic quantities due to the appearance of a factor $(sT+\mu_B\rho_B)$ in the denominator of eq.(\ref{s1pa}). However, this shifting does not occur at a fixed value of the chemical potential, rather moves slightly towards higher $\mu_B$ value with increasing temperature as shown in fig.(\ref{c1}b). For finite field strength $|eB|=0.04\ \text{GeV}^2$, as plotted in fig.(\ref{c2}), similar behavior of both the parallel and perpendicular components $c^{2(||,\perp)}_{s/\rho_B}$ are observed at smaller and higher values of $\mu_B$. Hence, the physical effects of crossings basically represent the non-monotonic behavior in the sound speed with respect to the variation in temperature and baryon chemical potential (density) of nuclear matter due to a shifting from the temperature driven influence to the density driven effect as given in eqs.(\ref{s1pa}-\ref{s1pen}) for the parallel and perpendicular components of the squared speed of sound along the isentropic curve. In fig.(\ref{c3}a), the magnitude of $c^{2\ (||,\perp )}_{s/\rho_B}$ increases with increasing temperature from
$T=10$ MeV to $50$ MeV, at low baryon density $\rho_B=\rho_0$, for $|eB|=0.04\ \text{GeV}^2$, a convenient choice for the current study of nuclear matter properties. However, at a comparatively high baryon density $\rho_B=4\rho_0$, $c^{2\ (||,\perp)}_{s/\rho_B}$ decreases as a function of temperature from
$T=18$ MeV to $50$ MeV. At low temperature region, the variation is observed to be non-monotonous.}
%The physical effect of crossing is more clearly visible with a plot of $c^2_{s/\rho_B}$ as a function of $T$ for different values of the nuclear matter density $\rho_B=\rho_0, 2\rho_0, 3\rho_0, 4\rho_0, 5\rho_0, 6\rho_0$ for a fixed field strength $|eB|=0.04GeV^2$, as shown in figs.(\ref{c3}). As the baryon density increases from $\rho_B=\rho_0- 6\rho_0$, the position of crossing in $c^2_{s/\rho_B}$ plot is shifted towards the high temperature regime.}

\section{Summary}
\label{sec5}
In summary, we have investigated the effective nucleon mass and various thermodynamic quantities in the magnetized symmetric nuclear matter at finite temperature and density. The pressure, magnetization, isothermal compressibility and the squared speed of sound are calculated from the thermodynamic potential of a grand canonical ensemble constructed within a scale-invariance breaking effective model Lagrangian with non-linear scalar self-interactions. The variation of the effective nucleon mass ($m^*_N$) with baryon chemical potential ($\mu_B$) shows that $m^*_N$ decreases abruptly within a specific range of $\mu_B$ indicating a transition from the vacuum to nuclear matter phase for different low temperatures at zero and finite magnetic field strengths. A monotonic rise in the effective nucleon mass is observed with increasing magnetic field, indicating the phenomenon of magnetic catalysis due to the magnetized vacuum contribution with zero anomalous magnetic moments of the nucleons. However, at large chemical potential, when the effective mass decreases considerably from its vacuum value, the behavior observed is slightly non-monotonous at low temperature. The thermodynamic pressure shows anisotropy with respect to the magnetic field direction and splits into two components along ($P^{||}$) and transverse ($P^{\perp}$) to the field direction. The amount of splitting increases with increasing magnetic field with negative values for the parallel component. The scaled magnetization exhibits oscillatory behavior for higher chemical potential and larger values of $|eB|$ for $T=10$ MeV due to the Landau level quantization of the charged fermions present in the magnetized nuclear matter. The oscillation disappears at high temperature. The isothermal compressibility which expresses the compression in system's volume with increasing pressure, exhibits the anisotropic nature of pressure separating into the components parallel ($K_T^{||}$) and perpendicular ($K_T^{\perp}$) with respect to the magnetic field direction. The parallel component is smaller than the perpendicular one indicating a stiffer equation of state along the direction of magnetic field. Next, the squared sound speed is shown to exceed the conformal limit of $1/3$ at higher chemical potential for both zero and finite magnetic field cases. The magnitude increases with rising temperature, which, however flips at large chemical potential and the transition point shifts towards lower chemical potential for low temperature. The observed in-medium behavior of the thermodynamic variables as well as effective nucleon mass are consistent with those obtained in the framework of non-linear Walecka model. However, the effects of the scalar gluonium field through the scale-invariance breaking logarithmic potential is seen to be marginal. In this work, the effects of QCD trace-anomaly condition is incorporated in the framework of hadronic effective Lagrangian and the effects on the thermodynamic variables are studied for a density regime suitable for neutron star matter study as well as in the low energy heavy ion collisions experiments, where matter up to several orders of nuclear saturation density are expected to be produced. The presence of a background magnetic field leads to the important phenomenon of magnetic catalysis while taking into account the effects of magnetized vacuum contribution. The field induced anisotropic structure in the thermodynamic variables of pressure, compressibility and speed of sound may provide significant insights into the characteristic properties of EOS of magnetized neutron star matter, suitable for the chosen range of medium parameters ($\mu_B,\ T$). %However, considerable field effects are observed for higher values of $|eB|$, which is somewhat not realized in the hadronic phase of heavy-ion collisions or neutron star matter. Still, the induced anisotropic structure may shed important light into the structural properties of dense matter system. 
%$\delta,\ \rho$
 
%\acknowledgements

\end{document}